\documentclass{amsart}
\usepackage{graphics,verbatim,color,amsthm}

\usepackage[all]{xy}

\theoremstyle{plain}
\newtheorem{theorem}{Theorem}

\newtheorem{cor}{Corollary}

\theoremstyle{remark}

\theoremstyle{definition}

\def\Ri{ Riemannian }

\def\R{\mathbb{R}}

\def\C{\mathbb{C}}

\def\H{\mathbb{H}}

\newcount\equationcount
\def\thenumber{0}
\def\eq#1{\global\advance\equationcount by 1
   \def\thenumber{\number\equationcount}
                        {$$#1\eqno(\thenumber)$$}}

\begin{document}

\title[MICZ]{MICZ-Kepler = dynamics  on   the cone over the rotation group}
\author{Richard Montgomery}
 
\address{Dept. of Mathematics\\ University of California, Santa Cruz\\ Santa Cruz CA}

\email{rmont@ucsc.edu}

\date{November 15, 2011 (Preliminary Version)}



\begin{abstract} We show that the n-dimensional MICZ-Kepler system arises from
 symplectic reduction of a simple  mechanical system
on the cone over the rotation group $SO(n)$.   As a corollary   we derive
an elementary formula for its general solution.  The punch-line of our computation
is that the additional  MICZ-Kepler $|\phi|^2/r^2$ type potential term
  is the rotational part of the cone's kinetic
energy. 
\end{abstract}

\maketitle


\section{Introduction} The classical mechanical  formulation  of a Hydrogen atom
is identical to the Kepler problem  for a   planet moving in the
gravitational field of a massive Sun.   
The MICZ [McIntosh-Cisneros-Zwanziger] -Kepler system is an integrable  extension of the Kepler
problem in which the charged proton at the origin of the Hydrogen atom  is simultaneously a   Dirac monopole.
The original references are \cite{Zwanziger, Cisneros}.  
See \cite{Meng1, Meng2} for   history,     references,  and  the n-dimensional generalization.     
Our purpose is to show that  the MICZ-Kepler  system is the reduction of a natural mechanical system
(no magnetic fields)  on the cone over the rotation
group.   See theorem 1 below and eq (\ref{H1}).  
We then use this realization  to   write down  an  explicit Lie-theoretic formula
(eq \ref{solution}) for the system's general solution.  

Meng \cite{Meng2}   formulated the MICZ-Kepler system as a system on what he called a  `Sternberg phase space'
-and what we will call the `adjoint bundle phase space' -
associated to  a  principal $SO(n-1)$ bundle over $\R^n \setminus \{0\}$.    
This  phase
space arises as  the symplectic reduction of  the   cotangent bundles of 
the same principal bundle.  See  \cite{{Montgomery1},{Montgomery2}, {Weinstein}}.
It follows that there is a Hamiltonian  system 
whose configuration space is   Meng's principal bundle  and whose reduction yields the MICZ-Kepler systems.
Our object is to find this system, and then use it. The heart of the computation
(noticing that the Hamiltonian   eq(\ref{H1})   becomes   (\ref{H2}) in adjoint bundle variables) 
 is the observation that the angular part of the Kepler kinetic energy plus
the quadratic color charge term of the MICZ-Keple Hamiltonian equal   the
kinetic energy for the bi-invariant metric on the full rotation group.  
(For the quantum-mechanical analogue of this computation
compare the algebra around  eq (\ref{H2}) to  the algebra at the end of section 2 of \cite{Zwanziger}. )
 
\section{MICZ-Kepler}  The MICZ -Kepler  system of equations can be written 
as a   mixed 1st-2nd order system for a curve $(q(t), \phi (t))$ in a vector bundle  over $\R^n \setminus \{0\}$.
Here $q(t)$ denotes the curve in $\R^n$ and   $\phi(t)$ lies in the  moving fiber   over this curve. 
This fiber is the Lie algebra ${\mathfrak so(n-1)}$  of $SO(n-1)$.  The bundle is called the   adjoint bundle,
and is an associated vector bundle to a certain  principal bundle  $SO(n-1) \to Q \to \R^n \setminus \{0\}$,
called the Dirac monopole and described in the next section.   So, relative to a local trivialization of the bundle, $\phi(t)$ takes values in the   Lie algebra $so(n-1)$.  The principal bundle 
is endowed with a canonical connection  whose curvature is   $F$, and which induces a covariant derivative
 $D$ on the adjoint bundle.  Then the MICZ-Kepler system is
$$\ddot q = - \frac{q}{r^3} + \frac{ |\phi |^2 q} {r^4} + \phi \cdot F( \dot q, \cdot)$$
$$D \phi / dt = 0.$$
Some clarifications are in order.  The curvature is a two-form with values in the adjoint bundle.
Consequently $F(\dot q, \cdot)$ is a one-form with values in the adjoint bundle.
The Killing form   endows the adjoint bundle with a natural fiber inner product,
so that $\phi \cdot F(, \dot q, \cdot)$ is a one-form on $\R^n$.  We turn this one-form into a vector using the
standard flat metric. In coordinates then: $ \phi \cdot F( \dot q, \cdot) ^j  = \phi_a F^a _{ij} (q) \dot q ^i $.
 In a local trivialization the second equation reads $D \phi/dt = d \phi /dt + [ A _i \dot q^i , \phi]$ where $A$
 is the connection one-form, an   ${\mathfrak so(n-1)}$-valued one-form , in this trivialization.

The system has   two basic conserved quantities: its energy
\begin{equation}
\label{E1}
H = \frac {1}{2}( | \dot q | ^2 + \frac{| \phi| ^2 }{r^2}) - \frac{1}{r},
\end{equation}
and the square norm of the adjoint variable, or $so(n-1)$- 
$$\text{Casimir} = \| \phi (t) \|^2.$$
There are other conserved quantities, an angular momentum and a Runge-Lenz (or Laplace) 
vector, but  we will not need them here.
 
The adjoint bundle  canonically fibers into adjoint-orbit fibers which are preserved by  parallel transport. Consequently  the fiber variable  $\phi (t)$ stays
on whichever adjoint orbit fiber it begins on at time $0$.   
As a particular case, the zero-orbit  is preserved and  corresponds to  setting $\phi = 0$ in the
equations,  which reduces them  to the  standard equations of the Kepler problem. 
Meng \cite{Meng2} takes $\phi$ to lie in an  adjoint orbit of  a particular type which 
he calls  ``magnetic''.  We   allow any $\phi$, hence
any  adjoint orbit.

\section{The n-dimensional Dirac Monopole}  
We describe the bundle-with-connection needed to define the MICZ-Kepler equations of  the previous section.
We take our description  from   Meng, who called it the Dirac monopole,  since that is what it is when $n = 3$.
 
  Consider the space $C_0$ of all orthogonal frames
$f = (f_1, f_2, \ldots, f_n)$ on $\R^n$  normalized so that their    lengths are all equal:
$|f_i| = |f_j|$ all $i,j$. 
The map 
$$f \mapsto f_n , \qquad  C_0 \to  \R^n \setminus \{0\},$$ 
  gives $C_0$ the structure of a principal   $SO(n-1)$ bundle over $\R^n \setminus \{0\}$. 
  Write $r =|f_n|$.  Then, we have
an $SO(n)$-equivariant diffeomorphism 
$$\R^+ \times SO(n) \to C_0 ;  (r, g) \mapsto (r ge_1, rg e_2, \ldots r ge_n) = (f_1, \ldots f_n),$$
where $e_1, \dots, e_n$ is the standard basis of $\R^n$.  From this perspective,
the bundle projection becomes   $(r, g) \mapsto r g e_n$.  

Put spherical coordinates on $\R^n$ so that $\R^n \setminus \{0\} \cong \R^+ \times S^{n-1}$.
Recall that $S^{n-1} = SO(n)/SO(n-1)$ where the $SO(n-1) \subset SO(n)$ is the stabilizer of
$e_n$.     The restriction of
the principal bundle $C_0 \to \R^n \setminus \{0\}$ to $S^{n-1} \subset \R^n \setminus \{0\}$
defines the homogeneous principal bundle $\pi_{S^{n-1}} : SO(n) \to S^{n-1}$.
The Killing form on $so(n)$ endows $SO(n)$ with a bi-invariant metric $d^2 s_{SO(n)}$, and relative to this
metric, the orthogonal complement to the fibers of the projection $\pi_{S^{n-1}}$
define an $SO(n)$-equivariant  connection  for the homogeneous principal bundle.
Extend this connection trivially in the radial ($r$) direction to arrive at
the ``Dirac connection'' of $C_0 \to \R^n \setminus \{0 \}$.    We refer to the bundle $C_0$
endowed with this connection as `the monopole'.    For coordinate expressions
for the connection  see \cite{{Zwanziger}, {Cisneros}, {Meng1},{Meng2}} .

\section{Adjoint and co-adjoint bundles.}  Suppose that  $G \to Q \to S$ is a principal $G$
bundle.  Then the adjoint bundle $Ad(Q) : = Q \times_G {\mathfrak g} \to S$
is the   vector bundle associated  to $Q$ via the  adjoint action of $G$ on ${\mathfrak g}$.
This means we divide the product $Q \times {\mathfrak g}$   by the $G$-induced equivalence relation
$(q, \xi) \sim (qg, Ad_{g ^{-1}} \xi)$.  Write equivalence classes $[q, \xi]$. 
The fiber of the adjoint bundle is ${\mathfrak g}$.

The co-adjoint bundle is defined similarly,   using the co-adjoint action.
If ${\mathfrak g}$ is endowed with a   bi-invariant inner product
such as the Killing form on ${\mathfrak g} = so(n-1)$, then we get $G$-equivariant isomorphisms
${\mathfrak g} \cong {\mathfrak g}^*$ and hence we can identify the adjoint bundle with
the co-adjoint bundle. This identification sends adjoint orbit fiber to the corresponding
co-adjoint orbit fiber.   We make this identification throughout the paper.
Thus the variable $\phi(t)$ in the MICZ-Kepler system is a section of the adjoint bundle
$Ad(C_0)$ along the curve $q(t) \in \R^n \setminus \{0\}$ 

A connection on $Q$ induces on any associated vector bundle, and so on the adjoint bundle.
We can describe parallel transport in ${\mathfrak g} (Q)$ along a curve $c(t) \in S$ as follows.
Pick a point $\phi (0) = [q(0), \xi] \in {\mathfrak g} (Q)$.  Consider the horizontal lift
$q(t) \in Q$ of $c(t)$.  Then $\phi(t) = [q(t), \xi] \in {\mathfrak g} (Q) $ is the parallel transport of $\phi(0)$
along $c(t)$. The   2nd MICZ-Kepler equation states that $\phi(t)$ is covariantly constant, and hence of
the form just described.  

\section{Metric cones and \Ri submersions}

Let $r = |f_n| \to 0$ in the  construction of $C_0 \cong \R^+ \times SO(n)$ so that 
  the whole of $SO(n)$
is crunched to a point defined by $r=0$.  We
we have formed the    cone  $C = Cone(SO(n)) \supset C_0$ over the full rotation group $SO(n)$.   
The bundle projection $C_0 \to \R^n \setminus \{0\}$ extends to the cone point $r=0$,
sending it to the origin.    

We now put a canonical metric on   the cone $C$.  

Recall that the  cone over a \Ri manifold $(X, d^2 s_X)$
is given by  the metric $dr^2 + r^2 d^2 s_X$ 
on $\R^+ \times X$ where $r \in \R^+$.   As $r \to 0$ the metric factor involving $X$ shrinks
to $0$ so it makes sense to  identify $\{0\} \times X$ to a single point, which is   the cone point.
In this we get a metric on $Cone(X) = [0, \infty) \times X/ \{0 \} \times X$
which is Riemannian away from the cone point.

We form the metric cone $C = Cone(SO(n))$ by 
applying this cone  construction to $X = SO(n)$ endowed with  a  Killing induced bi-invariant metric
$d^2 s _SO(n)$.   Such a metric is well-defined   up to scale.    That scale
will be fixed by insisting that the   bundle projection $C_0  \to \R^n \setminus \{0\}$ is  a \Ri submersion.

Recall the notion of a \Ri   submersion  $\pi: Y \to S$ between \Ri manifolds $Y, S$. 
Suppose that $\pi$ is a submersion.  Take any point $y \in Y$
and consider the orthogonal complement at $y$ to the  fiber  $\pi ^{-1} (s)$ through $y$. Here $s = \pi(y)$.
Following the principal bundle language as above, call this orthogonal complement ${\mathcal H}_y \subset T_y Y$
the `horizontal space' at $y$.   The differential $d \pi_y$ of $\pi$, restricted to the horizontal
space, is  necessarily a  linear bijection   onto the tangent space to $s$. Now the horizontal space inherits
an inner product from $Y$.   If this restricted differential is an isometry between inner product spaces
for all $y$,  then $\pi$ is said to be a \Ri submersion.     

Consider the standard metric $d^2 s_{S^{n-1}}$ on the unit sphere $S^{n-1}$.  
 The Killing scale for $SO(n)$ is now   fixed by insisting that the bundle projection   $SO(n) \to S^{n-1}$
 be  a \Ri submersion. This  scaling is the one
for which  the standard basis elements $e_i \wedge e_j$ of $so(n)$ have length $1$.  
Here $e_i \wedge e_j$ is the skew-symmetric operator sending $e_i$ to $e_j$ and
$e_j$ to $-e_i$, $i \ne j$.  Let $\theta_{ij}$ be the dual basis,
viewed as left-invariant one-forms.  Then
$d^2 s _{SO(n) } = \Sigma (\theta_{i j})^2$so that
\begin{eqnarray}
\label{metricC}
d^2 s _{C } &=&dr ^2 + r^2  \Sigma (\theta_{i j})^2  
\end{eqnarray}

We verify that $SO(n) \to S^{n-1}$ and $C_0 \to \R^n \setminus \{0\}$
are  \Ri submersions.  The connection form for the
Dirac monopole is the   $so(n-1)$-valued one-form 
\begin{equation}
\label{monopole}
A = \Sigma_{i < j <  n} \theta_{ij} e_i \wedge e_j , \qquad ( \text{ on } SO(n) \text{ or }  C_0 ).
\end{equation}
(Note we must use {\it left} invariant one forms, rather than right-invariant formsas can be seen
by the fact that the connection is not $G$-invariant, but rather $G$-equivariant
with $G$ acting on the Lie lagebra by the adjoint action.)
The horizontal distribution is defined by $\theta_{ij} =0,  i , j \ne n$.
The vertical distribution is defined by $\theta_{i n} = 0, i = 1, 2, \ldots , n-1$,
together with $dr = 0$ in the $C_0$ case.  Thus 
the Killing metric on $SO(n)$  splits  orthogonally relative to the horizontal-vertical splitting
\begin{eqnarray}
\label{metricS1}
d^2 s _{SO(n) } &=& \Sigma (\theta_{i n})^2 +  \Sigma_{ i < j < n}  (\theta_{ij})^2 \\
& = &  \pi^* d^2 s _{S^{n-1}} + d^2 s_{fiber}
\end{eqnarray}
The fact that $exp (t e_i\wedge e_n)$ has period $2 \pi$ shows that the metric  scalings are 
correct for the \Ri submersion:  a 2$\pi$ periodic horizontal geodesic in $SO(n)$ maps to a great circle of
circumference $2 \pi$ in $S^{n-1}$.  

Now, written out in spherical  coordinates the metric on $\R^n$ is
$ds^2 _{\R^n} = dr^2 + r^2 ds^2_{S^{n-1}}$, which is to say that  metrically speaking $\R^n = Cone(S^{n-1})$.
The corresponding radial-horizontal spherical-vertical splitting of the metric on $C$ is
\begin{eqnarray}
\label{metricS1}
d^2 s _{C } &=& (dr^2 + r^2  \Sigma (\theta_{i n})^2 )+  r^2 \Sigma_{ i < j < n}  (\theta_{ij})^2 \\
& = &  \pi^* ds^2 _{\R^n} + d^2 s_{fiber}
\end{eqnarray}
which shows    the projection $C \to \R^n$ becomes a \Ri submersion away from the cone point.    

\section{Main result: MICZ-Kepler from a mechanical system on the cone.}

A natural mechanical system consists of  a configuration space $Q$,
endowed with a \Ri metric $ds^2_Q$ and a potential function $V: Q \to \R$. 
This data defines a Hamiltonian on $T^*Q$ whose Hamiltonian $H$ is kinetic plus potential:
$H = K + V$, where the kinetic energy $K$ is induced by the metric.  In standard canonical coordinates 
$(q, p) = (q_i, p^i)$ for $T^*Q$ we have $K(q,p) = \frac{1}{2} g^{ij} (q) p_i p_i p_j$
if $ds^2 _Q = g_{ij} dq^i dq^j$.  

\begin{theorem} Take the canonical metric cone $C = Cone(SO(n))$ described in the 
previous section so as  to get a \Ri metric on $C_0 = Cone(SO(n) \setminus \{0\}) \to \R^n \setminus \{0 \}$.
Take potential function $V = -\frac{1}{r} :C_0 \to \R$  
where $r: Cone(SO(n)) \to [0, \infty)$ denote the cone's radial coordinate.
Then the resulting natural mechanical system (Hamiltonian (\ref{H1}) below)  on $T^* C_0$
is invariant under the lifted  action of $SO(n)$, and so is  also invariant
by $SO(n-1)$.  The symplectic  reduction of this system by $SO(n-1)$ at any particular $\mu \in so(n-1)^* \cong so(n-1)$
yields the  generalized MICZ-Kepler system associated to the adjoint orbit through
$\mu$. 
 \end{theorem} 
 
We see from the expression (\ref{metricC}) for the metric on the cone that the
Hamiltonian  of this theorem is
\begin{equation}
\label{H1}
H = \frac{1}{2} (p_r ^2 + \frac{1}{r^2} \Sigma \xi_{ij}^2 )- \frac{1}{r} 
 \end{equation}
 where $r,p_r$ are canonical coordinates on $T^* \R^+$
 and the $\xi_{ij}$ are the Lie-Poisson coordinates -- linear coordinates on $so(n)^*$ -
 induced by the choice of basis $e_i \wedge e_j$ for $so(n)$.
 
 \begin{cor}  Any solution $(q(t), \phi(t))$  to the generalized  MICZ-Kepler system 
 can be  constructed as follows.    Fix   $\xi \in so(n), w \in S^{n-1}$.  
Fix a solution $r(t)$ to the   1-dimensional Kepler problem:
 $\ddot r = - V_{\mu} ^{\prime} (r)$ where  the effective potential
 is $V_{\mu} (r)= -\frac{1}{r} + \frac{\mu^2}{ r^2}$
  with     $\mu^2 = | \xi |^2$, and 
   a  solution $u(t)$ to the ODE   $\dot u = \frac{1}{r(t)^2}$. 
 Then 
 \begin{equation} 
 \label{solution}
 q (t) = r(t) exp(u (t) \xi) w  
 \end{equation}
The adjoint bundle variable $\phi(t)$    is obtained  by parallel translating 
 an initial  adjoint vector $\phi(0) = [f, A_f (\xi)]$  along  $q(t)$, where $f \in C_0$ is any element  projecting to $q(0)$
 and $A$ is the connection one-form (\ref{monopole}).
 \end{cor}
 

 {\bf Special Cases.}
 
 {\bf  1. Kepler.}  Take $\xi$ horizontal over the initial $q(0)$, so 
 that $A(\xi) = 0$ and $\phi(t) = 0$.  For simplicity, 
 take  the initial $q(0)$ in the direction 
 $w = e_n$.  Then horizontality implies  $\xi = \Sigma v_i e_i \wedge e_n  = {\vec v} \wedge e_n$
 is an infinitesimal  rotation in the ${\vec v}, e_n$ plane.  The solution $q(t)$ then  lies in this plane.
 Set $\theta(t)  = u(t) | \xi| $.  Then  $(r(t),  \theta(t))$ is a solution to    Kepler's equations 
 expressed in polar coordinates. 
 
 {\bf 2. Magnetic cone.}  Meng takes his $\phi (0)$ to be of ``magnetic type',
 which means, relative to a local trivialization,  that $\phi(0) ^2 = - \mu Id.$
 In other words, up to scale $\phi(t)$ is an almost complex structure on $\R^n$,
 compatible to the standard complex structure.  Thus $\xi$  satisfies $\xi ^2 = -\mu ^2 Id$.
 Set $J = \xi/\mu$ so that $J$ is an honest almost complex structure.  
 Then   $exp(u  \xi ) =  cos( \mu u) I  + \sin (\mu u) J $.
 The solution $q(t)$ lies  on the  two-plane spanned by $q(0)$ and $J q(0)$.
   Indeed, it is another Keplerian conic on that plane, as Meng showed  in  \cite{Meng2}. 
 
{\bf  3. Generic.}  Take $\xi$ generic, meaning that  it has   $[n/2]$ distinct nonzero eigenvalues $\pm i \omega_j$
  linearly independent over the rationals.  We can, by conjugating by a  rotation,
put $\xi$ into the normal form $\Sigma \omega_j e_{2j -1} \wedge e_{2j}$.
Then  $\theta \mapsto exp(\theta \xi)$ is a dense curve on a standard maximal torus in
$SO(n)$.  For   negative energy the corresponding
   one dimensional Kepler motion
is periodic with  period $T$ , and without collision (since $\xi \ne 0$).  We  can arrange that $2 \pi /T$
is rationally independent of the $\omega_j$.
  Then the corresponding  solution curve $q(t)$ forms a dense winding
  on a kind of ``annular projection'' $r_{min} \le r \le r_{max}$ to $\R^n$ of a torus of dimension $[n/2] + 1$.

 \section{Proof of theorem 1.}  We apply the   general theory of reduction
 of    cotangent bundles of a principal bundle. We first describe that general theory.   See
   \cite{Montgomery1}, 
particularly pp. 160-163, or  the earlier references
  \cite{ Montgomery1, Weinstein, Sternberg} for perhaps more leisurely descriptions. 
 
 Let   $G \to Q \to S$ be a principal $G$-bundle.    $G$ acts on $T^*Q$ with
  $G$-equivariant
 momentum map $J: T^* Q \to {\mathfrak g}^*$.   The quotient 
 $(T^*Q)/G$ is  naturally a Poisson manifold whose symplectic leaves are
the symplectic  reduced spaces $J^{-1} ({\mathcal O}_ \mu)/G = J^{-1} (\mu)/G_{\mu}$
where ${\mathcal O}_{\mu}= G \cdot \mu$ denotes the coadjoint orbit through $\mu \in {\mathfrak g}^*$.
The general theory proceeds by using a connection on $Q$ to  
 define a symplectic isomorphism with these reduced  spaces.
 
  Differentiate the
  sequence of maps $\xymatrix{ G \ar[r] & Q  \ar[r]^{\pi} & S}$  at  fixed $q \in Q$ to 
  obtain the   sequence of   linear maps 
   $\xymatrix{  {\mathfrak g}  \ar[r]^{\sigma_q } &  TQ  \ar[r]^{d \pi_q} &   T_s S}$
   where   $s = \pi(q)$.   The sequence is exact: the  image of $\sigma_q$ equals the kernel of $d\pi_q$.
   (This common image is  called the vertical space at $q$.)    
    Letting   $q$  vary parametrically we   obtain the `Atiyah sequence' (described  in  \cite{Atiyah})
    \begin{equation}
    \label{AtiyahSeq}
       \xymatrix{ Q \times {\mathfrak g}  \ar[r]^{\sigma} &  TQ  \ar[r]^{d \pi} &  \pi_S ^* TS}
       \end{equation}
       which is  an exact
sequence of  $G$-equivariant vector bundles over $Q$.
    Dualizing the first map of (\ref{AtiyahSeq}), and composing with the   projection   
       yields the momentum map
      : $$J: \xymatrix{ T^*Q  \ar[r]^{\sigma^{*}} & Q \times {\mathfrak g}^{*} \ar[r]&  {\mathfrak g}^*}.$$
      
      A  connection   $A$ for $Q \to S$ induces a    $G$-invariant splitting of 
      (\ref{AtiyahSeq}) and hence a
        $G$-equivariant  isomorphism:  
$$TQ \cong \pi_S^* TS \oplus (Q \times {\mathfrak g}).$$
 Dualizing yields the   $G$-equivariant  isomorphism:  
\begin{equation}
T ^*Q \cong \pi_S^* T^* S \oplus (Q \times {\mathfrak g}^*).
\label{dualIso}
\end{equation}
Now $G$ acts on the bundle  $Q \times {\mathfrak g}$ by $g(q, \xi) = (qg, Ad_{g^-1} \xi)$ as per the equivalence relation  used to define   the  adjoint bundle 
and thus the action  of $G$ on $Q \times {\mathfrak g}^*$ is the one used to define the co-adjoint bundle. 
Forming the quotient by $G$ we thus get the bundle isomorphism
$$\Psi_A: (T^*Q)/G \cong T^*S \oplus Ad^* (Q)$$ over  $S$
which is our desired identification.    We refer to the right hand side
of this isomorphism as being ``on the Adjoint bundle side'' in what follows.

From our factorization of   $J$   we see that  
under the  isomorphism $\Psi_A$    the reduced spaces $J^{-1} (\mu)/G_{\mu} = J^{-1} ({\mathcal O}_ \mu)/G$
become the submanifolds $T^*S \oplus ({\mathcal O}_{\mu}) (Q)$, which are the Adjoint bundle phase spaces
and the Sternberg phase spaces of (\cite{Meng2}). 
If  ${\mathfrak g}$ is endowed with a bi-invariant Killing form as above, then  the co-adjoint orbit bundle
  $({\mathcal O}_{\mu}) (Q)$ is identified with a corresponding adjoint orbit. 


 We   need more  detail regarding  the isomorphism $\Psi_A$ and the Poisson brackets on the universal phase
  spaces in order to pull-back the cone Hamiltonian (\ref{H1}) and  compute  equations of motion.
    The connection defines    horizontal lift operators $h_q:  T_s S \to T_q Q$, $q \in Q$,
    which are linear operators  whose
    image is the horizontal space ${\mathcal H}_q = ker(A(q)) \subset T_q Q$ of the splitting of $TQ$.
    The dual of $h_q$ is $h_q ^*: T^* _q Q \to T^* _s S $ and is one factor of the
    isomorphism (\ref{dualIso}).   Write
    $[q, P]$ for the equivalence class in $T^*Q/G$ of $(q, P) \in T^*Q$.
    Then 
    $$\Psi_A ([q,P] ) = (\pi(q),  h_q^* P) \oplus [q,  J(q,P)]  \in T^*S \oplus Ad^* (Q).$$
 
   We describe $\Psi_A$  in coordinates.    Let $(x^i , g) \in \R^d \times G$ be coordinates
   on $Q$ induced by a   local trivialization $Q_U \cong U \times G$ 
   of $Q \to S$,  together with 
coordinates on $U \subset S$.   Then,  over $U$ we have $T^*Q \cong T^* U \times T^* G = T^*U \times G \times {\mathfrak g}^*$ 
with coordinates $(x^i, p_i,  g,  \xi_a)$
where the coordinates $\xi_a$ are Lie-Poisson linear coordinates on ${\frak g}^*$. 
relative to a basis $e_a$  of ${\frak g}$.   Thus $(T^* Q)/G \cong T^*U \times {\mathfrak g}^*$
with coordinates $x_i, p_i, \xi_a$. 
On the other hand,  the same   data $(x^i, g)$ and basis $e_a$ 
 yield   coordinates $x_i, \pi_i,  \xi_a$
for $T^* S \oplus Ad^* (Q)$.  Relative to these two sets of coordinates the map $\Psi_A$
is the minimal coupling procedure 
$\Psi_A (x, p_i, \xi) = (x^i , p_i - \xi_a A^a _i (x),  \xi_a) = (x^i , \pi_i, \xi_a)$
 where $A(x) = \Sigma e_a A^a _i (x) dx$ is the
connection one-form relative to the  local trivialization and coordinates. 
The brackets on the Adjoint bundle  side are
$\{x_i , \pi_j \} = \delta ^i _ j,  \{ \xi_a \xi_b \} = -c^d _{a b} \xi _d$
and $\{ \pi_i, \pi_j \} = -\xi_a F ^a_{ij}$,
$\{\pi_i, \xi_a \} = D_i  \xi ^a = [A_i, \xi]_a$.
Here $F^a _{ij}$ is the expression for the curvature of $A$
in this local trivialization.  (Compare eqs (12.2) of \cite{Montgomery1} to  (3.2) \cite{Meng2}.
Note that in  the triple of displayed equations immediately following (12.2) of
\cite{Montgomery1}   most  
terms should have a   capital $P$ immediately in front of them.) 
 These agree with the brackets found in Meng for the case of $so(n-1)$.

Now we recompute the Hamiltonian (\ref{H1})  on the Adjoint bundle side
using $\Psi_A$.  The dual of the metric splitting (\ref{metricS1}) yields
\begin{eqnarray}
\label{K1}
K_C & = &\frac{1}{2} [p_r ^2 + \frac{1}{r^2}(\Sigma_{i =1} ^n   \xi_{in} ^2 +  \frac{1}{2} \Sigma_{ i < j < n}  \xi_{ij} ^2)] \\
        & = & \frac{1}{2} [ p_r ^2 + \frac{1}{r^2} (h^*K_{S^n} + |\phi|^2) ]
\end{eqnarray}
where $h^* : T^* SO(n) \to T^* S^{n-1}$ is the connection induced
dual of the horizontal lift.
The   fiber term  $\Sigma_{ i < j < n}  \xi_{ij} ^2$ corresponds, on the Adjoint bundle side, 
to    the Casimir function $|\phi \|^2$
of $SO(n-1)$, viewed as a function on the adjoint bundle.   
So  the   Hamiltonian, viewed  on adjoint bundle side, 
reads 
$$H = \frac{1}{2} (p_r ^2 + \frac{1}{r^2} K_{S^n} +   \frac{1}{r^2} |\phi |^2)  - \frac{1}{r}$$
The sum of the first two terms  $ \frac{1}{2} (p_r ^2 + \frac{1}{r^2} K_{S^n})$
 is     the usual  kinetic energy  $\frac{1}{2} \Sigma_{i =1} ^n  \pi_i ^2 $
 on $\R^n$ written   in spherical variables.  Thus  
\begin{equation}
H = \frac{1}{2} (\Sigma \pi_i ^2 +   \frac{1}{r^2} |\phi |^2) - \frac{1}{r}
\label{H2}
\end{equation}
 which  is the  MICZ-Kepler Hamiltonian.

  \section{Proof of the corollary.}
  
 We   compute the equations of motion on $C_0$,  using the expression (\ref{H1}) for the
 Hamiltonian and the  fact that $\Omega = \|\xi \|^2$ is a Casimir. 
 Set $\mu^2 = \Omega$ and 
 $V_{\mu} (r) = -\frac{1}{r} + \frac{\mu^2}{r^2}$
 Then Hamilton's equations on $T^* C_0 = T^* \R^+  \times T^* SO(n) = \R^+ \times \R \times SO(n) \times so(n)$ 
 are: 
 $$\dot r = \{ r, H \} = p_r $$
 $$\dot p_r = \{p_r , H \} = - V_{\mu^2} ^{\prime} (r) $$
 $$\dot g = g \frac{ \partial H}{\partial \xi}  = g \frac{1}{r^2} \xi$$
 $$\dot \xi_a = \{\xi_a , H \} =\frac{1}{2r^2} \{ \xi_a, \Omega\} = 0  $$
 The first pair of  equations decouple from the second pair, and assert that  $(r,p_r)$ evolves as per the 
 one-dimensional radial Kepler equation with effective potential $V_{\mu}$.
 The last equation asserts that $\xi \in so(n)$ is constant.
 The equation for $g$ asserts
 that $g(t) = g_0 exp(u(t) \xi)$ where $du/dt = 1/r^2$.    Indeed, the solution to
 $\dot g = g \xi$ through $g = Id$ is  the  one-parameter subgroup $exp(t \xi)$ 
 and this flow   is generated by the Hamiltonian $\frac{1}{2} \Omega$.  We   have scaled the Hamiltonian
 on $SO(n)$  by the (time-dependent) factor $1/r^2$ and used left-invariance. 
To  rewrite $g_0 exp (u(t) \xi) = exp(u(t) \tilde \xi)g_0$ we can   
  set  $\tilde \xi = g_0 \xi g_0 ^{-1}$.

  We now have the  solution the Kepler equation on the cone:
  $(r(t), g(t))$ with $g(t) = exp(u(t) \xi) g_0$.  Recall the bundle projection
  is $(r, g) \mapsto r ge_n$ and use that any unit vector $w$  can be
  written $g_0 e_n$ to the expression for $q(t)$ in the corollary.    
  QED
  
  \section{Other groups}  The   tricks used here
  apply to  any Lie group $G$ in place of   $SO(n)$
  provided that  $G$ is endowed with  an  faithful orthogonal representation on $\R^n$
  which is   transitive on the unit sphere $S^{n-1}$.
  We get   the theorem that the  Kepler problem on 
  $Cone(G)$ is equivalent to the `MICZ-Kepler-$G$ ' problem whose
  `color variables' lie in an adjoint orbit bundle for $G$ over $\R^n \setminus \{0\}$. 
  The standard families of such groups are the unitary groups $U(n)$ 
  on $\R^{2n} = \C^n$ and $Sp(n; \H)$ on $\H^n = \R^{4n}$.

  \bibliographystyle{amsplain}

\end{document}